\documentclass[11pt, a4paper]{article}
\usepackage{jheparxiv}
\usepackage[latin1]{inputenc}
\usepackage{amsmath}
\usepackage{amsfonts}
\usepackage{amssymb}
\usepackage{latexsym}
\usepackage{mathrsfs}
\usepackage{graphicx}
\usepackage{color}
\usepackage{slashed}
\usepackage{twistor}
\usepackage[all]{xy}

\renewcommand{\d}{\mathrm{d}}
\renewcommand{\th}{\mathtt{h}}

\subheader{\hfill \texttt{DAMTP-2015-43}}

\title{Gravity with a cosmological constant from rational curves}

\author{Tim Adamo}

\affiliation{Department of Applied Mathematics \& Theoretical Physics \\
        University of Cambridge \\
        Wilberforce Road \\
        Cambridge CB3 0WA, United Kingdom}

\emailAdd{t.adamo@damtp.cam.ac.uk}

\abstract{We give a new formula for all tree-level correlators of boundary field insertions in gauged $\cN=8$ supergravity in AdS$_4$; this is an analogue of the tree-level S-matrix in anti-de Sitter space. The formula is written in terms of rational maps from the Riemann sphere to twistor space, with no reference to bulk perturbation theory. It is polynomial in the cosmological constant, and equal to the classical scattering amplitudes of supergravity in the flat space limit. The formula is manifestly supersymmetric, independent of gauge choices on twistor space, and equivalent to expressions computed via perturbation theory at 3-point $\overline{\mbox{MHV}}$ and $n$-point MHV. We also show that the formula factorizes and obeys BCFW recursion in twistor space.}

\notoc 

\begin{document}
 
\maketitle

\vfill

\pagebreak

\section{Introduction}

The scattering amplitudes of gravity possess many structures which are obscured by traditional approaches to their calculation based on perturbation theory of the Einstein-Hilbert action. At tree-level, a potent example of this is provided by a strikingly compact formula for the entire classical S-matrix of gravitons in any dimension~\cite{Cachazo:2013hca}. This formula is underpinned by the `scattering equations'~\cite{Fairlie:1972,Fairlie:2008dg,Gross:1987kza,Gross:1987ar} and based upon integrals over the moduli space of a marked Riemann sphere rather than any Feynman diagram expansion on space-time. 

In four dimensions, even greater simplifications are possible thanks to the structure of on-shell superspace. Indeed, all tree amplitudes of $\cN=8$ supergravity~\cite{Cachazo:2012kg} (as well as amplitudes of $\cN=4$ supersymmetric Yang-Mills and Einstein-Yang-Mills theory~\cite{Roiban:2004yf,Adamo:2015gia}) in four dimensions have compact, manifestly supersymmetric expressions when written in terms of rational maps from a marked Riemann sphere to \emph{twistor space}.\footnote{We remind the reader that twistor space $\PT$ is a suitably chosen open subspace of the complex projective space $\CP^{3|\cN}$. Each point $(x,\theta)$ in (complexified) chiral Minkowski superspace corresponds to a linearly embedded Riemann sphere in $X\subset\PT$. See~\cite{Penrose:1986ca,Ward:1990,Adamo:2013cra} for reviews.} The degree of the map fixes the MHV degree of the scattering amplitude: a degree $d$ map leads to a N$^{d-1}$MHV tree amplitude. 

These twistor expressions are best thought of as integral kernels for scattering amplitudes: to obtain explicit S-matrix elements, specific representatives for external states and a contour in the moduli space of rational maps must be chosen. When standard momentum eigenstates (written on twistor space) are used, the moduli integrals for the map can be performed explicitly, leading to delta functions enforcing a refinement of the scattering equations written in the spinor helicity formalism of four dimensions~\cite{Witten:2004cp}.


\medskip

It is well known that in space-times with a cosmological constant $\Lambda$, the gravitational S-matrix is replaced by a different observable whose definition depends on the sign of $\Lambda$. For anti-de Sitter (AdS) space, scattering amplitudes are replaced by correlators between field insertions on the time-like boundary which are propagated though the bulk by the space-time action via Witten diagrams. The AdS/CFT correspondence gives a non-perturbative definition of these objects in terms of correlation functions between appropriate local operators in the boundary CFT~\cite{Maldacena:1997re,Gubser:1998bc,Witten:1998qj}. For example, the analogue of a tree-level graviton scattering amplitude in AdS should be equal to the strong coupling limit of a correlation function of stress-energy tensors in the boundary CFT. In the flat space limit ($\Lambda\rightarrow0$) these graviton correlators should become scattering amplitudes of gravity in Minkowski space~\cite{Susskind:1998vk,Polchinski:1999ry,Giddings:1999qu,Giddings:1999jq}.\footnote{In de Sitter space ($\Lambda>0$), the situation is less canonical. While there is a mathematically computable S-matrix propagating asymptotic data from past to future infinity, no physical observer can measure it. Physical observables can be defined by restricting to the observable region of de Sitter space, but this may spoil gauge invariance (\textit{c.f.}, \cite{Witten:2001kn,Strominger:2001pn,Maldacena:2002vr,Maldacena:2011nz}).}   

While calculating explicit correlators -- even at tree-level -- with Witten diagrams is complicated, it is known that these objects share many properties with flat space scattering amplitudes. Particularly useful tools in this regard are the embedding space formalism and Mellin transform, which render Witten diagrams into a form closely resembling momentum space Feynman diagrams~\cite{Penedones:2010ue,Fitzpatrick:2011ia,Paulos:2011ie,Nandan:2011wc}. Furthermore, tree-level gravity correlators in AdS exhibit factorization behavior~\cite{Fitzpatrick:2011dm,Goncalves:2014rfa} and obey recursion relations~\cite{Raju:2010by,Raju:2011mp,Raju:2012zr} which are a natural generalization of BCFW recursion in flat space~\cite{Britto:2005fq,Benincasa:2007qj}.

\medskip

Despite their similarities, it is safe to say that nothing close to the compact expressions for gravity amplitudes in flat space has been found for tree level graviton correlators in AdS. In Mellin space, concrete results are often limited to external scalars in the bulk theory~\cite{Fitzpatrick:2011ia}, while recursive techniques have only yielded results at four points in gravity~\cite{Raju:2012zs}. While the computation of Witten diagrams can be substantially simplified using a conformal partial wave decomposition~\cite{Hijano:2015zsa}, this still operates diagram-by-diagram in perturbation theory. Yet if the surprisingly compact expressions for the S-matrix in flat space are a reflection of some deeper structure in the field theory itself, one should expect analogous expressions to exist regardless of the background. 

This paper provides evidence that this is true in one particular context: gauged $\cN=8$ supergravity in AdS$_4$. We propose a formula for (an integral kernel for) tree-level correlators in this theory based upon rational maps from a marked Riemann sphere to twistor space. In the bulk theory, this should correspond to field strength insertions on the AdS boundary with specified polarizations, propagated through the bulk by the classical supergravity action. In the pure gravity sector, the (anti-)self duality of these insertions gives a notion of MHV degree for such correlators. Supersymmetrically, a N$^{d-1}$MHV correlator will be homogeneous of degree $8(d+1)$ in the Grassmann variables of the $\cN=8$ supermultiplet insertion. A degree $d$ rational map to twistor space corresponds to a N$^{d-1}$MHV tree level correlator. 

The primary ingredients of the formula are: four arrays of differential operators acting on the external states; a choice of structure on twistor space -- an \emph{infinity twistor} -- whose role is to break conformal invariance by encoding the cosmological constant (as well as the coupling for the gauged $R$-symmetry); and a choice of `gauge' on twistor space amounting to $d+1$ arbitrary points on the Riemann sphere. 

Written in its most general form, this formula is a polynomial in $\Lambda$ whose $O(1)$ piece is the (integral kernel for the) flat space scattering amplitude. With a particularly auspicious choice of the `gauge' on twistor space, the formula becomes remarkably simple:
\be\label{intro1}
\cM_{n,d}=\int \d\mu_{n,d}\,\mathrm{det}'\!\left(\HH\right)\,\mathrm{det}'\!\left(\HH^{\vee}\right)\,\prod_{i=1}^{n} h_{i}(Z(\sigma_i))\,,
\ee    
where $\d\mu_{n,d}$ is the measure on the moduli space of maps from the Riemann sphere $\Sigma\cong\CP^1$ to twistor space, $Z:\Sigma\rightarrow\PT$; $\mathrm{det}'(\HH)$ and $\mathrm{det}'(\HH^{\vee})$ are reduced determinants of two $n\times n$ matrices depending on the infinity twistor and marked points $\{\sigma_i\}\subset\Sigma$; and $h_i(Z(\sigma_i))$ are the twistor wavefunctions. 

Schematically, \eqref{intro1} is equivalent to the Cachazo-Skinner formula for the tree-level S-matrix of $\cN=8$ supergravity~\cite{Cachazo:2012kg}, but there are two important differences. First, the matrices $\HH$, $\HH^{\vee}$ depend on the cosmological constant via the infinity twistor, making it difficult to integrate out the moduli in the same way as flat space and spoiling four-momentum conservation (as expected for correlators in AdS$_4$). Secondly, the structural equivalence holds only for a special choice of the twistor `gauge'; generally the formula contains many more terms (each of $O(\Lambda)$) which do not resemble the flat space amplitude. Even these general terms are remarkably compact, though.

\medskip

Of course, writing the formula in terms of rational curves in twistor space completely obscures its relationship with expressions for correlators obtained in the traditional way via Witten diagrams (or the strong coupling limit of some boundary correlation function). This integral kernel expression forestalls the choice of explicit boundary states -- which must be paired against the twistor wavefunctions to obtain expressions in position or momentum space -- as well as the selection of contour in the moduli space of rational maps appropriate to AdS$_4$. This is also true of twistor space expressions for the S-matrix in Minkowski space, but one expects that specifying the boundary states and explicitly integrating the moduli will be significantly more complicated for AdS.

Nevertheless, there are several powerful checks of the formula's validity which can be performed entirely at the level of the integral kernel in twistor space. These include checking twistor `gauge' invariance (a property with no obvious space-time interpretation that provides strong constraints in twistor space) and matching the formula with expressions equivalent to space-time perturbation theory for 3-point $\overline{\mbox{MHV}}$ and $n$-point MHV. Furthermore, the formula obeys BCFW recursion, whose structure is conformally invariant in twistor space~\cite{Mason:2009sa,ArkaniHamed:2009si}. While translating from the twistor recursion to those in Mellin~\cite{Fitzpatrick:2011ia} or momentum space~\cite{Raju:2010by,Raju:2012zr} may prove complicated, this establishes that our formula has the factorization properties expected for tree-level AdS correlators.



\section{The Formula}
\label{sec:Form}

Twistor space $\PT$ is an open subset of $\CP^{3|\cN}$ and can be charted with homogeneous coordinates
\be\label{tcoords}
Z^{I}=(Z^{a},\chi^{A})=(\mu^{\dot\alpha},\lambda_{\alpha},\chi^{A})\,,
\ee
for $a=1,\ldots,4$, $A=1,\ldots,\cN$. The superconformal group $\SL(4|\cN,\C)$ acts as linear transformations on these projective coordinates, so twistors are a natural set of variables for manifesting superconformal invariance. This fact underlies their utility in the study of $\cN=4$ super-Yang-Mills theory (\textit{c.f.}, \cite{Adamo:2011pv}).

Of course, to describe gravitational theories some additional structure is needed on $\PT$ to break superconformal invariance. This structure is known as the \emph{infinity twistor}~\cite{Penrose:1976js,Ward:1980am}. Just as the infinity twistor plays a crucial role when writing flat space scattering amplitudes in twistor space, we expect it to be equally central when writing gravity correlators for AdS$_4$. After briefly reviewing the role of the infinity twistor, we present our formula and explain its structure.


\subsection{Motivation}

From now on, we consider twistor space with $\cN=8$ extended supersymmetry. The infinity twistor is a (graded) skew bi-twistor $I_{IJ}$ which can be thought of as setting a mass scale. More specifically, if $X^{ab}=Z^{[a}_{1}Z^{b]}_{2}$ are homogeneous coordinates for the (bosonic) line $X\subset\PT$ corresponding to a space-time point $x$, then the infinity twistor defines the conformal factor for a conformally-flat space-time metric:
\be\label{met1}
\d s^{2}=\frac{\d X_{ab}\,\d X^{ab}}{(I_{cd} X^{cd})^2}\,.
\ee
Thus $I_{ab}$ gives the hypersurface `at infinity'; points which obey $I_{ab} X^{ab}=0$ live on the three-dimensional conformal boundary of space-time. The odd-odd components of the infinity twistor, $I_{AB}$, induce a metric on the $R$-symmetry group, so $I_{AB}\neq 0$ will correspond to gauging this $R$-symmetry~\cite{Wolf:2007tx}.

On (complexified) AdS$_4$, the simplest choice for infinity twistor compatible with the homogeneous coordinates \eqref{tcoords} is
\be\label{it1}
I_{IJ}=\left(\begin{array}{c|c}
				I_{ab}\,  & \,0 \\
				\hline
				0\,  & \,I_{AB}
	     \end{array}
\right)\,, \qquad 
I_{ab}=\begin{pmatrix}
	\Lambda \epsilon_{\dot\alpha\dot\beta} & 0 \\
	0 & \epsilon^{\alpha\beta}
       \end{pmatrix}\,, \qquad 
I_{AB}=\mathrm{g}\,\sqrt{\Lambda}\,\delta_{AB}\,.
\ee
Here $\Lambda$ is the cosmological constant of mass dimension $+2$ and $\mathrm{g}$ is a dimensionless coupling for the gauging of space-time $R$-symmetry. This particular choice for $I_{AB}$ breaks the $R$-symmetry group from $\SL(8,\C)$ to $\SO(8)$. Using standard incidence relations on $\PT$, the resulting metric is seen to be
\be\label{adsmet1}
\d s^2=\frac{\eta_{\mu\nu}\,\d x^{\mu}\,\d x^{\nu}}{(1+\Lambda x^2)^2}\,,
\ee
which is the AdS$_4$ metric written in an affine Minkowski space coordinate patch. Hence, our choice of infinity twistor is appropriate for describing gauged $\SO(8)$ supergravity on AdS$_4$.

So long as $\Lambda\neq0$ the infinity twistor is non-degenerate and has a well-defined inverse. We denote by $I^{IJ}$ the (graded) skew bi-twistor related to $I_{IJ}$ by $I^{IJ}I_{JK}=\Lambda\delta^{I}_{K}$. Given the choice \eqref{it1}, this `dual' infinity twistor is:
\be\label{dit}
I^{IJ}=\left(\begin{array}{cc|c}
				\epsilon^{\dot\alpha\dot\beta} & 0 & 0 \\
				0 & \Lambda\epsilon_{\alpha\beta} & 0 \\ 
				\hline
				\phantom{0} & \phantom{0} & \phantom{0} \\ [-1.2em]
				0 & 0 & \mathrm{g}^{-1} \sqrt \Lambda\, \delta^{AB}
			\end{array}
		\right)\, .
\ee
For short-hand, we denote contractions with the infinity twistor or its dual using angle or square brackets,
\begin{equation*}
 I_{IJ}\, A^{I}\,B^{J}=\left\la A,\,B\right\ra\,, \qquad I^{IJ}\,C_{I}\,D_{J}=\left[C,\,D\right]\,.
\end{equation*}
Geometrically, the infinity twistors define a weighted contact and Poisson structure on $\PT$, though we do not make use of this perspective here.

\medskip

In flat space, a $n$-point gravitational tree amplitude is proportional to $\kappa^{n-2}$, where $\kappa\sim (\mbox{mass})^{-1}$ is the gravitational coupling constant. Thus, when written in twistor space the amplitude expression must have $n-2$ insertions of the infinity twistors to balance the mass dimension~\cite{Cachazo:2012kg}. In AdS, the mass dimension of the tree correlators must still be $n-2$, but now overall powers of $\Lambda$ can contribute along with infinity twistor insertions. Simple arguments indicate that a N$^{d-1}$MHV tree correlator written in twistor space should be a monomial of degree $d$ in $\sqrt{\Lambda}$ and $\la\,,\,\ra$, and a monomial of degree $(n-d-2)$ in $\sqrt{\Lambda}$ and $[\,,\,]$. As desired, these numbers are exchanged by parity transformation.


\subsection{Tree correlators}

Our formula for the $n$-point, N$^{d-1}$MHV tree-level correlator of gauged $\cN=8$ supergravity in AdS$_4$ in the generic case where $n\gg d$ reads:
\begin{multline}\label{form1}
\cM_{n,d}=\int\frac{\prod_{r=0}^{d}\d^{4|8}\cZ_{r}}{\mathrm{vol}\;\GL(2,\C)}\left[\mathrm{det}'\!\left(\HH\right)\mathrm{det}'\!\left(\HH^{\vee}\right)+\Lambda\sum_{\substack{i \\ a,b}}\mathrm{det}'\!\left(\HH^{i}_{i}\right)\mathrm{det}'\!\left(\HH^{\vee\,a}_{\:\:\:\:b}\right)\,\mathfrak{h}^{ab}_{i} \right. \\
+\Lambda\sum_{\substack{i,j \\ a,b}} \mathrm{det}'\!\left(\HH^{ij}_{ij}\right)\mathrm{det}'\!\left(\HH^{\vee\,a}_{\:\:\:\:b}\right)\,\mathfrak{k}^{ab}_{ij}+\Lambda^2 \sum_{\substack{i,j \\ a,b,c,d}}\mathrm{det}'\!\left(\HH^{ij}_{ij}\right)\mathrm{det}'\!\left(\HH^{\vee\,ab}_{\:\:\:\:cd}\right) \mathfrak{h}^{ac}_{i}\, \mathfrak{h}^{bd}_{j}  \\
\left.+\cdots+\Lambda^{d}\sum_{\substack{i_1 ,\ldots,i_{2d} \\ a_{1},b_{1},\ldots,a_d,b_d}}\mathrm{det}'\!\left(\HH^{i_1 \cdots i_{2d}}_{i_1 \cdots i_{2d}}\right)\mathfrak{k}^{a_1 b_1}_{i_1 i_2}\cdots \mathfrak{k}^{a_d b_d}_{i_{2d-1} i_{2d}}\right] \prod_{i=1}^{n} h_{i}\!\left(Z(\sigma_i)\right)\,.
\end{multline}
Let us explain the various ingredients in this formula as well as its overall structure. The formula is based upon rational maps $Z:\Sigma\rightarrow\PT$; using homogeneous coordinates $\sigma^{\alpha}=(\sigma^{1},\sigma^{2})$ on $\Sigma\cong\CP^1$ we write such a map as a degree $d$ polynomial
\be\label{tmap}
Z^{I}(\sigma)=\sum_{r=0}^{d}\cZ_{r}^{I}\,(\sigma^{1})^{r} (\sigma^{2})^{d-r}\,.
\ee
The measure $\prod_{r=0}^{d}\d^{4|8}\cZ_r$ is over the coefficients of this map, and the quotient by $\mathrm{vol}\;\GL(2,\C)$ accounts for $\SL(2,\C)$ invariance on the rational curve and $\C^*$ scale invariance on $\PT$. The $\SL(2,\C)$-invariant inner product on $\Sigma$ is denoted $(i\,j)=\epsilon_{\alpha\beta}\sigma^{\alpha}_{i}\sigma^{\beta}_j$, and the weight $+2$ holomorphic measure by $\D\sigma=(\sigma\,\d\sigma)$.

Central to the formula are two $n\times n$ matrices, $\HH$ and $\HH^{\vee}$. The first of these is built from differential operators:
\be\label{H}
\HH_{ij}=\frac{\sqrt{\D\sigma_{i}\,\D\sigma_{j}}}{(i\,j)}\left[\frac{\partial}{\partial Z(\sigma_i)}, \frac{\partial}{\partial Z(\sigma_j)}\right]\,, \quad \mbox{for } i\neq j\,,
\ee 
\begin{equation*}
\HH_{ii}=-\sum_{j\neq i}\left[\frac{\partial}{\partial Z(\sigma_i)}, \frac{\partial}{\partial Z(\sigma_j)}\right]\,\frac{\D\sigma_i}{(i\,j)}\prod_{r=0}^{d}\frac{(p_{r}\,j)}{(p_{r}\,i)}\,,
\end{equation*}
while the second is algebraic with respect to the map to twistor space:
\be\label{Hv}
\HH^{\vee}_{ab}=\left\la Z(\sigma_a), Z(\sigma_b)\right\ra \frac{\sqrt{\D\sigma_{a}\,\D\sigma_{b}}}{(a\,b)}\,, \quad \mbox{for } a\neq b\,,
\ee
\begin{equation*}
\HH^{\vee}_{aa}=-\left\la Z(\sigma_a), \partial Z(\sigma_a)\right\ra\,.
\end{equation*}
The entries for both matrices are sensitive to the cosmological constant $\Lambda$ through the infinity twistors $I^{IJ}$ and $I_{IJ}$, respectively. The diagonal entries of $\HH$ depend on the choice of $d+1$ reference points $\{\sigma_{p_r}\}$ on $\Sigma$; we refer to this as a choice of `gauge' on twistor space.\footnote{More formally, this choice of `gauge' is equivalent to the choice of a kernel for the $\dbar$-operator acting on sections of $\cO(d)\rightarrow\CP^1$. The ambiguity of this choice is given by $H^0(\CP^1,\cO(d))$, which has dimension $d+1$.}

The quantities $\det'(\HH)$ and $\det'(\HH^\vee)$ are reduced determinants defined by removing rows and columns from $\HH$ and $\HH^{\vee}$ at the expense of a Jacobian factor. For $\det'(\HH)$, one removes $(d+2)$ rows and columns, while for $\det'(\HH^\vee)$ one removes $(n-d)$ rows and columns. Denoting the set of removed rows and columns (assumed to be identical for ease of notation) from $\HH$ and $\HH^{\vee}$ as $\th$, $\th^{\vee}$, respectively, the reduced determinants are given by:
\be\label{rds}
\mathrm{det}'\!\left(\HH\right):=\frac{\left|\HH^{\th}_{\th}\right|}{\prod_{i<j\in\th} (i\,j)^{2}}\prod_{k\in\th}\D\sigma_k\,, \qquad \mathrm{det}'\!\left(\HH^\vee \right):=\frac{\left|\HH^{\vee\,\th^{\vee}}_{\:\:\:\:\th^\vee}\right|}{\prod_{a<b\in\overline{\th^\vee}}(a\,b)^2} \prod_{c\in\overline{\th^\vee}}\D\sigma_{c}^{-1}
\ee  
where $\overline{\th^\vee}$ is the compliment of $\th^\vee$. Structurally, these reduced determinants are equivalent to those appearing in the Cachazo-Skinner formula for the flat-space S-matrix of $\cN=8$ supergravity~\cite{Cachazo:2012kg}. It can be shown that the formula is independent of the choice of rows and columns eliminated in $\det'(\HH)$, $\det'(\HH^{\vee})$; the proof is rather technical and will appear elsewhere. 

The remaining terms in the formula depend on two additional arrays of differential operators, denoted $\mathfrak{h}$ and $\mathfrak{k}$, whose entries are given by:
\be\label{fh}
\mathfrak{h}^{ab}_{i}=(-1)^{a+b}\frac{\D\sigma_{i} \sqrt{\D\sigma_{a}\D\sigma_{b}}}{(a\,b)}\left(\frac{Z^{I}(\sigma_{a})}{(i\,b)}\prod_{r=0}^{d}\frac{(p_r\,b)}{(p_r\,i)} - \frac{Z^{I}(\sigma_{b})}{(i\,a)}\prod_{r=0}^{d}\frac{(p_r\,a)}{(p_r\,i)}\right)\!\frac{\partial}{\partial Z^{I}\!(\sigma_i)}\,,
\ee
\begin{equation*}
\mathfrak{h}^{aa}_{i}=\frac{\D\sigma_i}{(i\,a)^2}\prod_{r=0}^{d}\frac{(p_r\,a)^2}{(p_r\,i)}\,\d_{a}\!\left(\frac{Z^{I}(\sigma_a)\,(i\,a)}{\prod_{s=0}^{d}(p_{s}\,a)}\right)\!\frac{\partial}{\partial Z^{I}\!(\sigma_i)}\,, \quad \mbox{for } i\in\overline{\th}\,, \; a,b\in\overline{\th^{\vee}}\,,
\end{equation*}
\be\label{fk}
\mathfrak{k}^{ab}_{ij}=(-1)^{a+b}\D\sigma_{i}\,\D\sigma_{j}\frac{\sqrt{\D\sigma_{a}\D\sigma_{b}}}{(a\,b)(i\,a)(j\,b)}\prod_{r=0}^{d}\frac{(p_{r}\,a)(p_r\,b)}{(p_{r}\,i) (p_{r}\,j)}\left[\frac{\partial}{\partial Z\!(\sigma_i)}, \frac{\partial}{\partial Z\!(\sigma_j)}\right]\,,
\ee
\begin{equation*}
\mathfrak{k}^{aa}_{ij}=\D\sigma_{i}\,\D\sigma_{j}\frac{\D\sigma_{a}\,(i\,j)}{(i\,a)^{2}\,(j\,a)^2}\prod_{r=0}^{d}\frac{(p_{r}\,a)^2}{(p_{r}\,i) (p_{r}\,j)}\left[\frac{\partial}{\partial Z(\sigma_i)}, \frac{\partial}{\partial Z(\sigma_j)}\right]\,, \quad \mbox{for } i,j\in\overline{\th}\,, \;a,b\in\overline{\th^{\vee}}\,.
\end{equation*}
These differential operators, as well as those in $\det'(\HH)$, act on the $n$ external states $h_i$, which are represented in terms of twistor wavefunctions.

The structure of \eqref{form1} is summarized as follows: each insertion of $\mathfrak{h}^{ab}_{i}$ removes an additional row and column ($i$) from $\det'(\HH)$ as well as an additional row ($a$) and column ($b$) from $\det'(\HH^{\vee})$, and then sums over all such choices. Likewise, each insertion of $\mathfrak{k}^{ab}_{ij}$ removes two additional rows and columns from $\det'(\HH)$ and a single row and column from $\det'(\HH^{\vee})$. So in \eqref{form1}, the various expressions hidden in $+\cdots+$ correspond to every way of combining increasingly many insertions of these arrays. At each stage, our notation for the summations is condensed, representing the symmetric sum over those indices not yet removed from the determinants.

As a concrete example, consider contributions to $\cM_{n,d}$ with an overall power of $\Lambda^2$. In our notation, these take the form
\begin{multline*}
\Lambda^{2}\!\int\!\frac{\prod_{r=0}^{d}\d^{4|8}\cZ_{r}}{\mathrm{vol}\;\GL(2,\C)}\left[\sum_{\substack{i,j \\ a,b,c,d}}\mathrm{det}'\!\left(\HH^{ij}_{ij}\right)\mathrm{det}'\!\left(\HH^{\vee\,ab}_{\:\:\:\:cd}\right)\, \mathfrak{h}^{ac}_{i}\, \mathfrak{h}^{bd}_{j} \right. \\
\left.+\!\!\!\sum_{\substack{i,j,k \\ a,b,c,d}}\!\!\mathrm{det}'\!\left(\HH^{ijk}_{ijk}\right)\mathrm{det}'\!\left(\HH^{\vee\,ab}_{\:\:\:\:cd}\right)\mathfrak{h}^{ac}_{i}\, \mathfrak{k}^{bd}_{jk}
+\!\! \sum_{\substack{i,j,k,l \\ a,b,c,d}}\!\!\!\mathrm{det}'\!\left(\HH^{ijkl}_{ijkl}\right)\mathrm{det}'\!\left(\HH^{\vee\,ab}_{\:\:\:\:cd}\right)\,\mathfrak{k}^{ac}_{ij}\, \mathfrak{k}^{bd}_{kl}\right]\! \prod_{i=1}^{n} h_{i}\!\left(Z(\sigma_i)\right)\,,
\end{multline*} 
where the summations are over
\begin{equation*}
\sum_{\substack{i,j \\ a,b,c,d}} = \sum_{\substack{i,j\in\overline{\th} \\ i\neq j}} \:\:\:\sum_{\substack{a,b,c,d\in\overline{\th^{\vee}} \\ a\neq b,\, c\neq d}}\,,
\end{equation*}
and so forth.

In the generic case $n\gg d$, this process terminates after exhausting $\det'(\HH^{\vee})$ by removing an additional $d$ rows and columns, with the final term containing $d$ insertions of $\mathfrak{k}$. If $(n-d-2)\leq d$, the sum of contributions to $\cM_{n,d}$ will terminate when $\det'(\HH)$ is exhausted instead. 

\medskip

External states are represented in this formula by insertions of the twistor wavefunction $h(Z(\sigma))$. For physical external states, these are appropriately chosen holomorphic $(0,1)$-forms on twistor space, homogeneous of degree $+2$ in the map $Z(\sigma)$; in other words, $h(Z(\sigma))\in H^{0,1}(\PT,\cO(2))$. Expanding in the fermionic coordinates $\chi^{A}$ of the map,
\be\label{es1}
h\!\left(Z^{a},\chi^{A}\right)=h(Z^a)+\chi^{A}\,\psi_{A}(Z^a)+\frac{1}{2}\chi^{A}\chi^{B} a_{AB}(Z^a)+\cdots+\chi^{8}\tilde{h}(Z^a)\,,
\ee
with each bosonic component related to a helicity sector of the $\cN=8$ supergravity multiplet by the Penrose transform~\cite{Penrose:1969ae,Eastwood:1981jy}. The precise choice of these physical states relevant for boundary insertions in AdS$_4$ is subtle, so in this paper we always assume that $h(Z(\sigma))$ is an `elemental state' on twistor space. In this way, the correlator \eqref{form1} is valued in $\Omega^{0,2n}(\bigoplus_{i=1}^{n}\PT_{i})$ and can be integrated against physical wavefunctions to obtain an expression on position or momentum space (see~\cite{Adamo:2011pv,Adamo:2013cra} for further discussion of this construction).

Two particular representations of such elemental states will be useful for us. \emph{Dual twistor} wavefunctions represent an external state as a plane wave specified by a choice of a dual twistor $W_i\in\PT^{\vee}$:
\be\label{dtwf}
h_{i}\!\left(Z(\sigma_i)\right)=\int \frac{\d t_{i}}{t_{i}^3}\,\e^{\im W_{i}\cdot Z(\sigma_i)}\,.
\ee
An obvious advantage of such wavefunctions is that they render the entries of $\HH$, $\mathfrak{h}$, and $\mathfrak{k}$ algebraic, replacing differential operators with dual twistors. The second useful representation is an eigenstate of a fixed point in twistor space; these wavefunctions enforce the projective coincidence of a fixed point and the rational map evaluated at the point $\sigma\in\Sigma$:
\be\label{eewf}
h_{i}\!\left(Z(\sigma_i)\right)=\bar{\delta}^{3|8}\!\left(Z_i,\,Z(\sigma_i)\right)=\int \frac{\d t_i}{t_i^{3}}\,\bar{\delta}^{4|8}\,\left(Z_{i}+t_{i} Z(\sigma_i)\right)\,.
\ee
Such wavefunctions are particularly useful when investigating BCFW recursion on twistor space.

\medskip

Our formula is a polynomial in $\Lambda$ of order $(n-2)$, whose $O(1)$ piece is given by the $\Lambda\rightarrow0$ limit of \eqref{form1}. When $\Lambda=0$, the infinity twistors become degenerate,
\begin{equation*}
\left[\frac{\partial}{\partial Z(\sigma_i)}, \frac{\partial}{\partial Z(\sigma_j)}\right]\rightarrow\left[\frac{\partial}{\partial \mu(\sigma_i)}, \frac{\partial}{\partial \mu(\sigma_j)}\right]\,, \qquad  \left\la Z(\sigma_a), Z(\sigma_b)\right\ra\rightarrow \left\la \lambda(\sigma_a), \lambda(\sigma_b)\right\ra\,,
\end{equation*}
and $\HH$, $\HH^{\vee}$ pass to matrices $\tilde{\Phi}$, $\Phi$ which appear in the Cachazo-Skinner formula for the tree-level S-matrix of $\cN=8$ supergravity~\cite{Cachazo:2012kg}. So keeping generic elemental states 
\be\label{flatlim1}
\cM_{n,d}\xrightarrow{\Lambda\rightarrow0}\int\frac{\prod_{r=0}^{d}\d^{4|8}\cZ_{r}}{\mathrm{vol}\;\GL(2,\C)}\,\mathrm{det}'\!(\tilde{\Phi})\,\mathrm{det}'\!(\Phi)\,\prod_{i=1}^{n}h_i\!\left(Z(\sigma_i)\right)\,,
\ee
and our formula for the tree-level correlator in AdS$_4$ passes smoothly to the S-matrix in the flat space limit.

It is easy to see that \eqref{form1} obeys the counting of infinity twistors and cosmological constants discussed above. In particular, $\det'(\HH)$ and $\det'(\HH^{\vee})$ contain $n-d-2$ powers of $[\,,\,]$ and $d$ powers of $\la\,,\,\ra$, respectively, so the leading term in $\cM_{n,d}$ has the same mass dimension counting as the flat space scattering amplitude. As further rows and columns are removed from $\det'(\HH)$ and $\det'(\HH^{\vee})$, we lose powers of the infinity twistors, but these are compensated with overall powers of the cosmological constant, $\Lambda$. From the definitions \eqref{fh}, \eqref{fk} it is easy to see that $\cM_{n,d}$ is -- term-by-term -- a monomial in $\sqrt{\Lambda}$, $[\,,\,]$ of degree $n-d-2$, and a monomial in $\sqrt{\Lambda}$, $\la\,,\,\ra$ of degree $d$.

\medskip

As a final observation, note that \eqref{form1} simplifies substantially with a particular choice of the `gauge' on twistor space. At this point, we have not yet demonstrated that this gauge (\textit{i.e.}, the $d+1$ points $\{\sigma_{p_r}\}$) can be freely chosen, though this is certainly the case if the formula is to be meaningful. In the following section, we check this `gauge' invariance explicitly, but for now suppose that it holds. This means that the $d+1$ reference points can take any values on $\Sigma$, so let $d$ of them coincide with the $d$ remaining rows and columns in $\det'(\HH^{\vee})$: $\{\sigma_{p_1},\ldots, \sigma_{p_d}\}=\overline{\th^{\vee}}$.

Then \eqref{fh}, \eqref{fk} ensure that each term involving entries from $\mathfrak{h}$, $\mathfrak{k}$ vanishes, since there is always a numerator factor of $\prod_{r=0}^{d}(p_r\,a)$. This renders the correlator's structure on twistor space equivalent to that of the S-matrix in flat space:
\be\label{gch0}
\cM_{n,d}=\int\frac{\prod_{r=0}^{d}\d^{4|8}\cZ_{r}}{\mathrm{vol}\;\GL(2,\C)}\,\mathrm{det}'\!\left(\HH\right)\mathrm{det}'\!\left(\HH^{\vee}\right) \prod_{i=1}^{n}h_i\!\left(Z(\sigma_i)\right)\,.
\ee
So if \eqref{form1} is `gauge' invariant, then the only distinctions between \eqref{gch0} and the (integral kernel of the) S-matrix are the non-degenerate infinity twistors associated with AdS$_4$ and the fixed gauge choice for $d$ of the reference points on $\Sigma$.


\section{Justification}
\label{sec:Evid}

Our formula is consistent with mass dimension counting (based on the infinity twistor and cosmological constant) and has the integral kernel of the tree-level S-matrix as its flat space limit. In this section we provide evidence which justifies the formula, establishing its `gauge' invariance on twistor space and correspondence with action-based calculations in the $\overline{\mbox{MHV}}$ and MHV sectors. Crucially, we demonstrate that the formula obeys BCFW recursion in twistor space.


\subsection{`Gauge' invariance on twistor space}

The expression for $\cM_{n,d}$ appears to have complicated dependence on the choice of $d+1$ points $\{\sigma_{p_r}\}$ on $\Sigma$, which we refer to as a choice of `gauge' on twistor space. Clearly, there is no space-time analogue for these points, so it must be the case that the formula is actually independent of them. This `gauge' also features in the Cachazo-Skinner expression for the flat space S-matrix, where `gauge' invariance follows from momentum conservation~\cite{Cachazo:2012kg}. Of course, that argument does not work in the AdS$_4$ context of interest here.

For simplicity, represent all $n$ external states by the dual twistor wavefunctions \eqref{dtwf}, and represent the rational map $Z^{I}(\sigma)$ as a degree $d$ polynomial
\be\label{gi1}
Z^{I}(\sigma)=\cZ^{I}_{\alpha_1\cdots\alpha_d}\sigma^{\alpha_1}\cdots\sigma^{\alpha_d}\,,
\ee
where the coefficients $\cZ^{I}_{\alpha_1\cdots\alpha_d}=\cZ^{I}_{(\alpha_1\cdots\alpha_d)}$ are the moduli of the map. It is also useful to introduce the shorthand
\be\label{gmom}
\cP_{I}^{\alpha_1\cdots\alpha_d}:=\sum_{i=1}^{n}t_i\,W_{i\,I}\,\sigma_{i}^{(\alpha_1}\cdots\sigma_{i}^{\alpha_d)}\,,
\ee
which can be thought of as a generalized momentum for the $n$-point correlator. With the choice of dual twistor wavefunctions, the entries of the matrix $\HH$ and the arrays $\mathfrak{h}$, $\mathfrak{k}$ become algebraic: all differential operators are replaced with insertions of dual twistors.


Without loss of generality, consider the dependence of $\cM_{n,d}$ on the reference point $\sigma_{p_0}:=\xi\in\Sigma$. We test this dependence by differentiating \eqref{form1} with respect to $\xi$; basic properties of determinants ensure that:
\begin{multline}\label{gi2}
 \d_{\xi}\cM_{n,d}=\int\frac{\d^{4|8(d+1)}\cZ}{\mathrm{vol}\;\GL(2,\C)}\left[\sum_{i}\mathrm{det}'\!\left(\HH^{i}_{i}\right)\mathrm{det}'\!\left(\HH^{\vee}\right) \d_{\xi}\HH_{ii}+\Lambda\sum_{\substack{i \\ a,b}}\mathrm{det}'\!\left(\HH^{i}_{i}\right)\mathrm{det}'\!\left(\HH^{\vee\,a}_{\:\:\:\:b}\right)\d_{\xi}\mathfrak{h}^{ab}_{i} \right. \\
+\Lambda\sum_{\substack{i,j \\ a,b}} \mathrm{det}'\!\left(\HH^{ij}_{ij}\right)\mathrm{det}'\!\left(\HH^{\vee\,a}_{\:\:\:\:b}\right)\,\d_{\xi}\HH_{jj}\,\mathfrak{h}^{ab}_{i}+\Lambda \sum_{\substack{i,j \\ a,b}}\mathrm{det}'\!\left(\HH^{ij}_{ij}\right)\mathrm{det}'\!\left(\HH^{\vee\,a}_{\:\:\:\:b}\right) \d_{\xi}\mathfrak{k}^{ab}_{ij}  \\
\left.+\cdots+\Lambda^{d}\!\!\!\sum_{\substack{i_1 ,\ldots,i_{2d},j \\ a_{1},b_{1},\ldots,a_{d},b_{d}}}\!\!\!\mathrm{det}'\!\left(\HH^{i_1 \cdots i_{2d}j}_{i_1 \cdots i_{2d}j}\right)\d_{\xi}\HH_{jj}\,\mathfrak{k}^{a_1 b_1}_{i_1 i_2}\cdots \mathfrak{k}^{a_d b_d}_{i_{2d-1} i_{2d}}\right] \e^{\im\cP\cdot\cZ}\prod_{i=1}^{n} \frac{\d t_i}{t_{i}^3}.
\end{multline}
A straightforward calculation reveals that the various matrix or array components depending on $\xi$ have derivatives
\be\label{xih}
\d_{\xi}\HH_{ii}=t_{i}\,\D\sigma_{i}\, \D\xi\, I^{IJ}W_{i\,I}\frac{\cP^{\alpha_{1}\cdots\alpha_d}_{J}}{(\xi\,i)^2}\prod_{r=1}^{d}\frac{\sigma_{p_r\,\alpha_r}}{(p_r\,i)}\,,
\ee
\be\label{xifh}
\d_{\xi}\mathfrak{h}^{ab}_{i}=\im\,t_{i}(-1)^{a+b}\D\sigma_{i}\frac{\D\xi\sqrt{\D\sigma_{a}\D\sigma_{b}}}{(a\,b)(\xi\,i)^2}W_{i\,I}\left(Z^{I}(\sigma_a)\prod_{r=1}^{d}\frac{(p_r\,b)}{(p_r\,i)} -Z^{I}(\sigma_b)\prod_{r=1}^{d}\frac{(p_r\,a)}{(p_r\,i)}\right)\,,
\ee
\begin{equation*}
\d_{\xi}\mathfrak{h}^{aa}_{i}=\im\,t_{i}\,\D\sigma_{i}\, \D\xi \frac{W_{i\,I}}{(\xi\,i)^2}\left(\partial Z^{I}(\sigma_{a})\prod_{r=1}^{d}\frac{(p_r\,a)}{(p_r\,i)} - Z^{I}(\sigma_a)\d_{a}\prod_{r=1}^{d}\frac{(p_r\,a)}{(p_r\,i)}\right)\,,
\end{equation*}
\begin{multline}\label{xifk}
\d_{\xi}\mathfrak{k}^{ab}_{ij}=-t_{i}t_{j}(-1)^{a+b}\frac{\D\sigma_{i}\D\sigma_{j}\sqrt{\D\sigma_{a}\D\sigma_{b}}\D\xi}{(a\,b)(i\,a)(j\,b)(\xi\,i)^2(\xi\,j)^2} \left((\xi\,b)(\xi\,j)(i\,a)+(\xi\,a)(\xi\,i)(j\,b)\right) \\
\times [W_{i},W_{j}]\,\prod_{r=1}^{d}\frac{(p_r\,a)(p_r\,b)}{(p_r\,i)(p_r\,j)}
\end{multline}
\begin{equation*}
\d_{\xi}\mathfrak{k}^{aa}_{ij}=-t_{i}t_{j}\D\sigma_{i} \D\sigma_{j} \D\sigma_{a} \D\xi \frac{[W_{i},W_{j}] (i\,j) (\xi\,a)}{(i\,a)^{2}(j\,a)^{2}(\xi\,i)^{2}(\xi\,j)^{2}} \left((\xi\,j)(i\,a)+(\xi\,i)(j\,a)\right)\prod_{r=1}^{d}\frac{(p_r\,a)^2}{(p_r\,i)(p_r\,j)} .
\end{equation*}
An important observation regarding these derivatives is that they are related to each other and to $\det'(\HH^{\vee})$ by derivatives with respect to the map moduli $\cZ^{I}_{\alpha_1\cdots\alpha_d}$.

In particular, a straightforward (if somewhat tedious) calculation reveals that:
\be\label{modrel1}
-\im\,t_{i}\,\D\sigma_{i}\,\D\xi\, \frac{I^{IJ}W_{i\,I}}{(\xi\,i)^2}\frac{\partial\,\mathrm{det}'\!\left(\HH^{\vee}\right)}{\partial\cZ^{J}_{\alpha_1\cdots\alpha_d}}\prod_{r=1}^{d}\frac{\sigma_{p_r\,\alpha_r}}{(p_r\,i)}=\Lambda \sum_{a,b}\mathrm{det}'\!\left(\HH^{\vee\,a}_{\:\:\:\:b}\right)\d_{\xi}\,\mathfrak{h}^{ab}_{i}\,,
\ee
\be\label{modrel2}
-\im\,t_{i}\,\D\sigma_{i}\,\D\xi\, \frac{I^{IJ}W_{i\,I}}{(\xi\,i)^2}\frac{\partial\,\mathfrak{h}^{ab}_{j}}{\partial\cZ^{J}_{\alpha_1\cdots\alpha_d}}\prod_{r=1}^{d}\frac{\sigma_{p_r\,\alpha_r}}{(p_r\,i)}=\d_{\xi}\,\mathfrak{k}^{ab}_{ij}\,.
\ee
As a consequence of these relations, the contributions to \eqref{gi2} can be grouped together based upon how many rows and columns have been removed from $\det'(\HH)$. For instance, all terms proportional to $\det'(\HH^{i}_{i})$ can be rewritten as the total derivative:
\begin{equation*}
 -\im\D\xi\frac{\partial}{\partial\cZ^{J}_{\alpha_1\cdots\alpha_d}}\left(\e^{\im\cP\cdot\cZ}\sum_{i}\mathrm{det}'\!\left(\HH^{i}_{i}\right)\mathrm{det}'\!\left(\HH^{\vee}\right)t_{i}\D\sigma_{i}\frac{I^{IJ}W_{i\,I}}{(\xi\,i)^2}\prod_{r=1}^{d}\frac{\sigma_{p_r\,\alpha_r}}{(p_r\,i)}\right)\,,
\end{equation*}
and likewise for all terms proportional to $\det'(\HH^{ij}_{ij})$
\begin{equation*}
-\im\D\xi\frac{\partial}{\partial\cZ^{J}_{\alpha_1\cdots\alpha_d}}\left(\e^{\im\cP\cdot\cZ}\sum_{\substack{i,j \\ a,b}}\mathrm{det}'\!\left(\HH^{ij}_{ij}\right)\mathrm{det}'\!\left(\HH^{\vee\,a}_{\:\:\:\:b}\right)\,\mathfrak{h}^{ab}_{i}\,t_{j}\D\sigma_{j}\frac{I^{IJ}W_{j\,I}}{(\xi\,j)^2}\prod_{r=1}^{d}\frac{\sigma_{p_r\,\alpha_r}}{(p_r\,j)}\right)\,.
\end{equation*}
Proceeding in this fashion, the derivative $\d_{\xi}\cM_{n,d}$ can be rewritten as
\begin{multline}\label{gi3}
 \d_{\xi}\cM_{n,d}=\int\frac{\d^{4|8(d+1)}\cZ}{\mathrm{vol}\;\GL(2,\C)}\,\left[\frac{\partial \cV^{I}_{\alpha_1\cdots\alpha_d}}{\partial\cZ^{I}_{\alpha_1\cdots\alpha_d}} \right. \\
 \left.+\Lambda^{d}\left(\prod_{i=1}^{n} \frac{\d t_i}{t_{i}^3}\right)\e^{\im\cP\cdot\cZ}\!\!\!\sum_{\substack{i_1 ,\ldots,i_{2d},j \\ a_{1},b_{1},\ldots,a_{d},b_{d}}}\!\!\!\mathrm{det}'\!\left(\HH^{i_1 \cdots i_{2d}j}_{i_1 \cdots i_{2d}j}\right)\d_{\xi}\HH_{jj}\,\mathfrak{k}^{a_1 b_1}_{i_1 i_2}\cdots \mathfrak{k}^{a_d b_d}_{i_{2d-1} i_{2d}}\right] \,,
\end{multline}
where $\cV^{J}_{\alpha_1\cdots\alpha_d}$ is smooth with respect to the map moduli.

The second line in \eqref{gi3} cannot be put into the form of a divergence, but vanishes after performing all integrals. Note that the only moduli dependence of this contribution is in the exponential $\e^{\im\cP\cdot\cZ}$, so we can do the moduli integrals to find
\begin{equation*}
 \int \frac{\Lambda^{d}}{\mathrm{vol}\;\GL(2,\C)} \delta^{4|8(d+1)}(\cP)\,\left(\prod_{i=1}^{n} \frac{\d t_i}{t_{i}^3}\right)\,\cP_{I}^{\alpha_1\cdots\alpha_d}\,\cF^{I}_{\alpha_1\cdots\alpha_d} = 0\,,
\end{equation*}
since $\cF$ is a function of the $\sigma_i$, $t_i$, and $W_i$ only. Therefore, the variation of $\cM_{n,d}$ with respect to the point $\xi\in\Sigma$ vanishes as a total derivative on the moduli space:
\be\label{gi4}
\d_{\xi}\cM_{n,d}=\int\frac{\d^{4|8(d+1)}\cZ}{\mathrm{vol}\;\GL(2,\C)}\,\frac{\partial \cV^{I}_{\alpha_1\cdots\alpha_d}}{\partial\cZ^{I}_{\alpha_1\cdots\alpha_d}}=0\,.
\ee
This establishes that our formula for the correlator is actually independent of the choice of $\{\sigma_{p_r}\}$, or `gauge' invariant on twistor space. Thus, we are always free to choose the `gauge' where the twistor space expression simplifies to \eqref{gch0}.


\subsection{$\overline{\mbox{MHV}}$ and MHV sectors}

The integral kernel of tree-level correlators for gauged $\cN=8$ supergravity in AdS$_4$ can also be obtained by classical perturbation theory in twistor space for the $\overline{\mbox{MHV}}$ and MHV sectors. These perturbative calculations are based on action functionals in twistor space which are (perturbatively) equivalent to the classical supergravity action in space-time, and hence provide an important check for our formula. 

In twistor space, the self-dual sector of supergravity is described by a holomorphic Chern-Simons action~\cite{Mason:2007ct}:
\be\label{sdta}
S[h]=\int_{\PT}\D^{3|8}Z\wedge h\wedge\left(\dbar h+\frac{1}{3}\left[\frac{\partial h}{\partial Z}, \frac{\partial h}{\partial Z}\right]\right)\,,
\ee
where $\D^{3|8}Z$ is the weight $-4$ holomorphic projective measure on twistor space, and $h\in\Omega^{0,1}(\PT,\cO(2))$ encodes the $\cN=8$ supergravity multiplet. This action is equivalent to the space-time theory in the sense that its equations of motion are equal to the equations of motion of gauged, self-dual $\cN=8$ supergravity on AdS$_4$~\cite{Wolf:2007tx,Mason:2007ct}.

Taking $h$ to be an elemental state on twistor space, the cubic vertex of this action provides the integral kernel for the $\overline{\mbox{MHV}}$ 3-point correlator:
\be\label{MHVbar}
\cM_{3,0}=\int_{\PT}\D^{3|8}Z\, h_{1}(Z)\, \left[\frac{\partial h_{2}(Z)}{\partial Z}, \frac{\partial h_{3}(Z)}{\partial Z}\right]\,.
\ee
Setting $n=3$, $d=0$ in \eqref{form1} leads to:
\begin{equation*}
\int \frac{\d^{4|8}\cZ}{\mathrm{vol}\;\GL(2,\C)}\frac{\D\sigma_1\D\sigma_2\D\sigma_3}{(1\,2)(2\,3)(3\,1)}\,h_{1}(\cZ)\,\left[\frac{\partial h_{2}(\cZ)}{\partial \cZ}, \frac{\partial h_{3}(\cZ)}{\partial \cZ}\right] = \cM_{3,0}\,,
\end{equation*}
with the equality following after using vol $\SL(2,\C)$ to eliminate dependence on $\sigma_1$, $\sigma_2$, $\sigma_3$, and vol $\C^*$ to projectivize $\d^{4|8}\cZ$ to $\D^{3|8}Z$.

\medskip

In twistor space, there is also an action functional describing conformal gravity and its (minimal) $\cN\leq4$ supersymmetric extensions~\cite{Mason:2005zm,Adamo:2013cra}, in the sense that solutions to the equations of motion on twistor space are in one-to-one correspondence with solutions of the equations of motion on space-time, up to diffeomorphisms. It is known that classical Einstein gravity on (Lorentzian) de Sitter space is equivalent to conformal gravity asymptotically restricted to Einstein degrees of freedom with an appropriate normalization~\cite{Maldacena:2011mk}. Analytic continuation implies that this statement holds for Euclidean or complexified AdS$_4$ as well.\footnote{Differing definitions of semi-classical observables in dS$_4$ or AdS$_4$ can be relegated to the choice of external states and contour in twistor space. Since the integral kernel is a polynomial in $\Lambda$, we do not expect its functional form to depend on the sign of $\Lambda$.}

The restriction to Einstein degrees of freedom inside conformal gravity is easily achieved on twistor space by means of the infinity twistor~\cite{Adamo:2012nn}. In~\cite{Adamo:2013tja}, it was shown how this could be used to obtain a generating functional for all MHV correlators on twistor space from the Einstein reduction of the conformal gravity action. The result is a formula for the $n$-point MHV correlator, written supersymmertrically as
\be\label{MHV1}
\cM_{n,1}=\frac{(n-3)!}{2\,n!\,\Lambda}\int\frac{\prod_{r=0,1}\d^{4|8}\cZ_{r}}{\mathrm{vol}\;\GL(2,\C)}\sum_{i,j} \frac{\left|\HH^{ij}_{ij}\right|}{(i\,j)^4}\,\HH^{\vee}_{ii}\,\HH^{\vee}_{jj}\,\prod_{k=1}^{n}h_{k}\!\left(Z(\sigma_k)\right)\,,
\ee
with a choice of `gauge' $\sigma_{p_1}=\sigma_i$. Via an integration-by-parts on the moduli space~\cite{Adamo:2013tja}, this expression is equal to
\begin{multline}\label{MHV2}
\cM_{n,1}=\int\frac{\prod_{r=0,1}\d^{4|8}\cZ_{r}}{\mathrm{vol}\;\GL(2,\C)}\frac{\left|\HH^{ijl}_{ijl}\right|\,\D\sigma_j\D\sigma_l}{(i\,j)^2 (j\,l)^2 (l\,i)^2}\,\HH^{\vee}_{ii}\,\prod_{k=1}^{n}h_{k}\!\left(Z(\sigma_k)\right) \\
=\int\frac{\prod_{r=0,1}\d^{4|8}\cZ_{r}}{\mathrm{vol}\;\GL(2,\C)} \mathrm{det}'(\HH)\,\mathrm{det}'(\HH^{\vee})\,\prod_{i=1}^{n}h_{i}\!\left(Z(\sigma_i)\right)\,,
\end{multline}
as desired.


\subsection{BCFW recursion}

It is a remarkable fact that the BCFW recursion relation is both simple and conformally invariant -- at the structural level -- when written in twistor space~\cite{Mason:2009sa,ArkaniHamed:2009si}:
\be\label{BCFW}
\cM_{n,d}(Z_1,\ldots,Z_n)=\sum_{\substack{n_L +n_R=n+2 \\ d_L +d_R=d}} \int \D^{3|8}Z\,\frac{\d q}{q}\,\cM_{n_L,d_L}(Z_1,\ldots,Z)\,\cM_{n_R,d_R}(Z,\ldots,Z_n-qZ_1)\,,
\ee
where the external states are assumed to be elemental on twistor space. The conformal invariance of gravity is `hidden' in the left and right subamplitudes of \eqref{BCFW}. So while the form of recursion relations for correlators in AdS$_4$ may differ from BCFW when written in momentum or Mellin space, there is no difference at the level of the integral kernel written in twistor variables.

In order to show that \eqref{form1} obeys this recursion relation, we must demonstrate that it has the correct large $q$ behavior under the BCFW shift, and that it factorizes on a simple pole in the moduli of the rational map to twistor space. In both regards, the formula demonstrates the desired properties in much the same fashion as the flat-space scattering amplitudes~\cite{Cachazo:2012pz}. 

In the first instance, consider \eqref{form1} for external states represented by dual twistor wavefunctions \eqref{dtwf}, with BCFW shift given by $W_{n}\rightarrow W_{n}-qW_{1}$. Without loss of generality, we choose to eliminate the rows and columns corresponding to $1,n$ in both $\det'(\HH)$ and $\det'(\HH^{\vee})$: $1, n\in\th, \th^{\vee}$. Further, we are free to choose a `gauge' with $p_{0}=n$, $p_{1}=1$. This ensures that the only dependence on $1$ and $n$ is via the external states and the Vandermonde determinant in the definition \eqref{rds} of $\det'(\HH)$.

In the affine coordinate patch $\sigma^{\alpha}=(1,z)$, external wavefunction insertions read:
\begin{equation*}
 \int \left(\prod_{i=1}^{n}\frac{\d t_{i}}{t_{i}^3}\right) \mathrm{exp}\left(\im\,W_{1\,I}\sum_{r=0}^{d}\cZ_{r}^{I}(t_1 z_1^r-qt_n z_n^r)+\im\sum_{j=2}^{n}t_{j}W_{j}\cdot Z(z_j)\right)\,.
\end{equation*}
Following~\cite{Cachazo:2012pz}, define a new scale parameter and affine coordinate by:
\be\label{scales}
\hat{t}_{1}:=t_{1}-qt_{n}\,, \qquad \hat{t}_{1}\,\hat{z}_{1}:=t_{1} z_{1}^{d}-q\,t_{n}z_{n}^{d}\,.
\ee
As the shift parameter $q$ becomes very large, the argument of the wavefunction exponential behaves as
\begin{equation*}
 \im\, W_{1}\cdot\cZ_{d}\,\hat{t}_{1}\hat{z}_{1}^{d}+\im\sum_{j=2}^{n}t_{j}W_{j}\cdot Z(z_j)+O\left(\frac{1}{q}\right)\,.
\end{equation*}
The only potential $q$-dependence of the correlator in the $q\rightarrow\infty$ limit is in the quantity
\be\label{falloff}
 \frac{\d t_{1}}{t_{1}^3}\,\frac{\D\sigma_1}{\prod_{j\in\th,\,j\neq1}(1\,j)}\xrightarrow{q\rightarrow\infty}\left(\frac{1}{q^3}\frac{\d\hat{t}_{1}}{t_{n}^3}\right)\left(\frac{\hat{t}_{1}}{q t_{n}}z_{n}^{\frac{d-1}{d}}\hat{z}_{1}^{d-1}\d\hat{z}_{1}\right)\left(\frac{q\,t_{n}}{z_{n}\hat{t}_{1}(1+\hat{z}_1^{d}/z_{n}^{d})}\right) \sim \frac{1}{q^2}
\ee
Not only is there no `pole at infinity' with respect to the BCFW shift, but we recover precisely the $q^{-2}$ falloff expected for gravity~\cite{ArkaniHamed:2008yf}.\footnote{Note that the Lagrangian-based techniques for observing this falloff are also applicable to the Einstein-Hilbert action with a cosmological constant (\textit{c.f.,} \cite{Adamo:2012nn}).}

\medskip

The factorization properties of our formula also follow in a manner practically identical to the flat-space calculation~\cite{Cachazo:2012pz}. To this end, it is useful to choose the `gauge' on twistor space with $\{\sigma_{p_1},\ldots\sigma_{p_d}\}=\overline{\th^{\vee}}$, whereupon the correlator takes the compact form \eqref{gch0}. Denote the remaining, un-fixed reference point by $\sigma_{p_0}=\sigma_*$. We model the factorization limit as a degeneration of the underlying rational curve, which can be represented as a quadric in $\CP^2$:
\begin{equation*}
 \{xy=s^2z\}\subset\CP^2\,.
\end{equation*}
In the $s\rightarrow0$ limit, this quadric degenerates into two rational curves $\Sigma_L$, $\Sigma_R$ joined at a node. If the degree of the non-degenerate map is $d$, then these two components are mapped to $\PT$ in the degenerate limit at degrees $d_L$ and $d_R$, respectively, with $d_L+d_R=d$. The affine coordinate $z$ on $\Sigma$ is related to the natural affine coordinates $u\in\Sigma_L$, $w\in\Sigma_R$ by
\be\label{affcoords}
u=\frac{s}{z}\,, \qquad w=sz\,.
\ee
The parameter $s^2$ serves as a coordinate transverse to the boundary divisor of the moduli space of rational maps represented by this degeneration.

From this point, one follows the same steps as in flat space (see~\cite{Cachazo:2012pz}) to deduce that
\begin{multline}\label{fact}
\cM_{n,d}=\int \D^{3|8}Z\,\frac{\d s^{2}}{s^2}\left(\frac{\prod_{s=0}^{d_L}\d^{4|8}\mathcal{U}_{s}}{\mathrm{vol}\;\GL(2,\C)}\, \mathrm{det}'\!(\HH_{L})\,\mathrm{det}'\!(\HH^{\vee}_{L})\, \prod_{i\in L\cup *}h_{i}(Z(u_i))\right. \\
\times\;\left. \frac{\prod_{t=0}^{d_R}\d^{4|8}\mathcal{W}_{t}}{\mathrm{vol}\;\GL(2,\C)}\,\mathrm{det}'\!(\HH_{R})\,\mathrm{det}'(\HH^{\vee}_{R})\,\prod_{k\in R\cup *}h_{k}(Z(w_k))\: +O(s^2)\right)\,,
\end{multline}
where $L,R$ are the sets of external states appearing on $\Sigma_{L},\Sigma_{R}$ in the $s\rightarrow0$ limit. The degree $d_L$, $d_R$ maps from each component of the degenerate curve have coefficients $\{\mathcal{U}_{s}\}$, $\{\mathcal{W}_{t}\}$, while the new states located at $\sigma_*$ on both $\Sigma_L$ and $\Sigma_R$ are represented by elemental states:
\begin{equation*}
 h_{*}(Z(u_*))=\bar{\delta}^{3|8}(Z, Z(u_*))\,, \qquad h_{*}(Z(w_*))=\bar{\delta}^{3|8}(Z, Z(w_*))\,. 
\end{equation*}
The entries of $\HH_{L}$, $\HH_{R}$ are appropriate for the degeneration, with the selection of reference points on $\Sigma_{L,R}$ given by $\{\sigma_{*}\}\cup\overline{\th^{\vee}_{L,R}}$. `Gauge' invariance on twistor space then implies that \eqref{fact} is actually equivalent to
\be\label{fact1}
\cM_{n,d}=\int \D^{3|8}Z\,\frac{\d s^{2}}{s^2}\left(\cM_{L}(Z_{i\in L}, Z)\,\cM_{R}(Z, Z_{k\in R})+O(s^2)\right)\,,
\ee
where $\cM_{L,R}$ are sub-correlators given by \eqref{form1}. 

Hence, our formula factorizes on a simple pole in the moduli space. The structural equivalence between the correlator and the S-matrix (in an appropriate `gauge') ensures that there are no other unphysical poles in the moduli space. Combined with the correct large $q$ falloff shown above, this suffices to prove that the correlator obeys the twistor space BCFW recursion relation \eqref{BCFW}. 

Note that unlike in flat-space, we cannot immediately identify the factorization pole in moduli with a momentum space propagator going on-shell. In flat space, momentum eigenstates can be inserted for the external particles and moduli integrals performed to obtain delta functions. These delta functions in turn imply that as the rational curve degenerates, the four-momentum exchanged between the two branches goes on-shell. In AdS$_4$, non-degenerate infinity twistors prevent straightforward integration of the moduli; this is a reflection of the fact that four-momentum is not conserved for correlators in AdS$_4$. So although our formula for the correlator factorizes in twistor space, it remains a non-trivial task to translate the BCFW recursion back to momentum (or Mellin) space for AdS$_4$.


\section{Discussion}

In this paper, we proposed a formula for all tree-level correlators of gauged $\cN=8$ supergravity in AdS$_4$ based on rational maps from the Riemann sphere to twistor space. This is a generalization of the Cachazo-Skinner formula for the tree-level S-matrix of $\cN=8$ supergravity~\cite{Cachazo:2012kg} to a setting with cosmological constant. We showed that this expression is well-defined and passes several non-trivial checks in its favor, including: consistent mass-dimension counting, a smooth flat space limit, matching with action-based perturbative calculations, and -- most importantly -- BCFW recursion in twistor space.

By treating the formula as an integral kernel for the supergravity correlator, most of the subtleties associated with semi-classical observables in AdS (\textit{i.e.}, boundary conditions, the precise form of asymptotic states, etc.) are relegated to the choice of external wavefunctions and a contour of integration in the moduli space. We avoided any discussion of these issues here, but understanding them in detail -- even for low numbers of external states -- seems an important next step. In particular, one would like to compare our formula (evaluated to an expression on position or momentum space) with `standard' calculations of supergravity correlators. The Penrose transform naturally corresponds to linearized field strengths or potentials on space-time rather than metric perturbations, so we expect our correlator is related to correlators of boundary stress tensor insertions by some integro-differential relation.

Expressing the S-matrix of a field theory in twistor space is related to re-writing its perturbation theory in terms of a \emph{twistor-string theory}~\cite{Witten:2003nn,Berkovits:2004hg,Skinner:2013xp,Engelund:2014sqa}. In the case of $\cN=8$ supergravity, this twistor-string should be able to incorporate the non-degenerate infinity twistor of AdS$_4$~\cite{Skinner:2013xp}. It would be interesting to see how the formula presented here emerges from the worldsheet perturbation theory of that model.

It is natural to wonder if compact formulae -- divorced from space-time perturbation theory -- exist for analogues of scattering amplitudes on other backgrounds or in higher dimensions. This is particularly true from the AdS/CFT perspective, since $\cN=8$ supergravity does not exist as any well-defined limit of string theory compactified to four dimensions~\cite{Green:2007zzb}. Formulae for the tree-level S-matrix of gravitons (in any number of dimensions) based on the scattering equations~\cite{Cachazo:2013hca,Mason:2013sva} are related to rewriting gravity at the non-linear level as a solvable 2d CFT~\cite{Adamo:2014wea}. This underlying simplicity seems a strong hint that further structure can be found in the semi-classical observables of gravity on any background space-time.

\acknowledgments

I would like to thank Eduardo Casali, Lionel Mason, Miguel Paulos, Jo\~ao Penedones, Eric Perlmutter, and David Skinner for useful comments and conversations. This work is supported by a Title A Research Fellowship at St. John's College, Cambridge.

\bibliography{AdS}
\bibliographystyle{JHEP}

\end{document}